\documentclass[conference]{IEEEtran}
\IEEEoverridecommandlockouts
\usepackage{cite}
\usepackage{amsmath,amssymb,amsfonts}
\usepackage{algorithmic}
\usepackage{graphicx}
\usepackage{textcomp}
\usepackage{xcolor}
\usepackage{url}
\usepackage{multirow}
\usepackage{tabularx}
\usepackage{tikz}
\usepackage{lipsum} 
\def\BibTeX{{\rm B\kern-.05em{\sc i\kern-.025em b}\kern-.08em
    T\kern-.1667em\lower.7ex\hbox{E}\kern-.125emX}}
\setkeys{Gin}{width=0.96\linewidth}
\begin{document}

\title{3D Point Cloud Object Detection on Edge Devices for Split Computing}

\author{
\IEEEauthorblockN{Taisuke Noguchi and Takuya Azumi}
\IEEEauthorblockA{\textit{Graduate School of Science and Engineering} \\
\textit{Saitama University}
}
}

\maketitle

\begin{abstract}
The field of autonomous driving technology is rapidly advancing, with deep learning being a key component. 
Particularly in the field of sensing, 3D point cloud data collected by LiDAR is utilized to run deep neural network models for 3D object detection.
However, these state-of-the-art models are complex, leading to longer processing times and increased power consumption on edge devices.
The objective of this study is to address these issues by leveraging Split Computing, a distributed machine learning inference method.
Split Computing aims to lessen the computational burden on edge devices, thereby reducing processing time and power consumption.
Furthermore, it minimizes the risk of data breaches by only transmitting intermediate data from the deep neural network model.
Experimental results show that splitting after voxelization reduces the inference time by 70.8\% and the edge device execution time by 90.0\%.
When splitting within the network, the inference time is reduced by up to 57.1\%, and the edge device execution time is reduced by up to 69.5\%.

\end{abstract}


\makeatletter
\def\ieeecopyright{ 
\footnotesize © 2024 IEEE. Personal use of this material is permitted.\newline DOI: 10.1109/RAGE62451.2024.00009
}
\makeatother 

\AddToHook{shipout/firstpage}{% 
\begin{tikzpicture}[remember picture,overlay] 
\node[anchor=south west,xshift=1.0cm,yshift=0.8cm] at (current page.south west)% 
{\parbox{\linewidth}{\raggedright\ieeecopyright}}; 
\end{tikzpicture}% 
}

\section{Introduction}
In recent years, the field of autonomous driving technology~\cite{autoware} has experienced rapid advancements, promising significant societal benefits.
These include the reduction of traffic congestion, a decrease in traffic accidents, and the introduction of innovative transportation options.
Central to this technology is object detection, a critical process for understanding the vehicle's environment, which is fundamental for effective route planning and vehicle control.
This detection primarily employs sensors installed in autonomous vehicles, complemented by infrastructure sensors on roads.
Infrastructure sensors can extend detection capabilities by covering areas that vehicle-mounted sensors cannot see, such as blind spots. 
Sharing this information with autonomous vehicles facilitates a more comprehensive and accurate object detection scope, greatly enhancing overall safety. 
Notably, LiDAR is a prominent sensor used in these systems, featured in both vehicle and infrastructure setups~\cite{roslite}.

LiDAR collects point cloud data of the surrounding environment and uses the data to estimate the position and classification of objects.
This data is crucial for object detection using \textit{Deep Learning} (DL).
In detail, the point cloud data is processed by a \textit{deep neural network} (DNN) model designed for 3D object detection~\cite{shi2020pv,deng2021voxel,guan2022m3detr} and executed.
In the context of infrastructure sensors, these DNN models operate on edge devices. 
However, this approach encounters challenges, particularly with the computational demands of advanced DNN models on edge devices. 
This leads to issues such as time-consuming processing, increased power consumption, and potential safety concerns due to less accurate lightweight models. 
To supplement the limited computing power of edge devices, alternatives should be considered.

Edge devices used for infrastructure sensor systems face a major challenge: state-of-the-art DNN models surpass the computational limits of edge devices.
Recent DNN models require significant power, leading to longer processing times and higher energy consumption, compromising real-time performance.
Lightweight models~\cite{li2023lightweight} offer a solution but at the cost of reduced accuracy and safety.
To address this problem, a method using edge servers with higher computational power has been proposed~\cite{edgecomputing}.
This involves transmitting point cloud data from edge devices directly to edge servers for processing.
However, this poses risks such as potential privacy information leakage due to data interception and communication delays due to the large size of LiDAR data.
These challenges are fatal, increasing processing time, decreasing accuracy, and leaking privacy information.
The objectives of this research are to reduce processing time, reduce the computational load on edge devices, and protect privacy.

This paper proposes a \textit{Split Computing} (SC)~\cite{kang2017neurosurgeon, SC_survey} method for 3D object detection using point cloud data.
In SC, the DNN model is split into two parts in the middle.
The first half of the model is executed on the edge device, and the output is transferred to the edge server, while the second half is executed on the edge server.
This method significantly reduces the amount of computation at the edge device, shortening processing time and reducing computational loads.
Furthermore, instead of sending the point cloud data obtained from LiDAR as is, the data generated in the middle of the DNN model are sent.
This reduces the amount of information in the sent data and improves privacy performance.

The contributions of this paper are as follows:
\begin{itemize}
\item
Reduction of time for 3D object detection in edge devices.
Allocating the majority of the object detection process to edge servers with higher computing power reduces the time required for object detection.
\item
Reduction of computational loads on edge devices in 3D object detection.
Executing the majority of the object detection process at edge servers reduces the amount of computation at edge devices.
\item
Performance comparison at multiple split points.
Comparison of the inference time and the edge device execution time at each split point to investigate the appropriate split point.
\end{itemize}

The remainder of this paper is organized as follows.
Section~\ref{system_model} provides a detailed description of SC and a toolbox for LiDAR-based 3D object detection in this study.
Section~\ref{approach} describes the proposed framework.
Section~\ref{evaluation} reports the evaluation of the proposed framework, and Section~\ref{related_work} discusses related work.
Finally, a brief conclusion is given in Section~\ref{conclusion}.

\section{SYSTEM MODEL} \label{system_model}

\begin{figure}[t]
        \centerline{\includegraphics{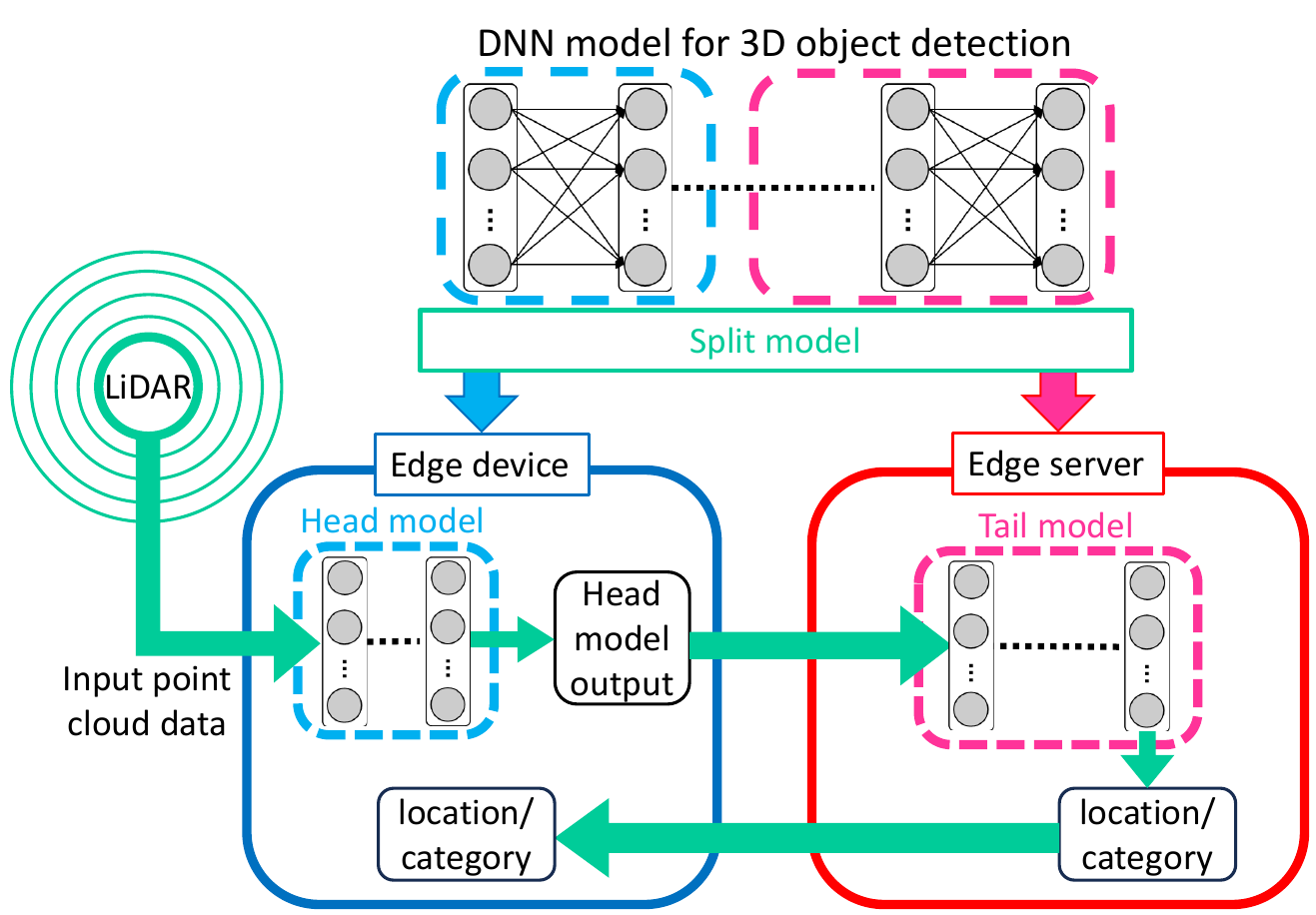}}  
        \caption{System Model.}
    \label{fig-system_model}
    \vspace{-2mm}
\end{figure}

The system model of this paper is shown in Fig.~\ref{fig-system_model}.
In this study, a method called Split Computing is used for 3D object detection at edge devices using point cloud data from LiDAR.
This section provides background on Split Computing and OpenPCDet.
Section~\ref{split_computing} and Section~\ref{openpcdet} describe Split Computing and OpenPCDet, respectively.

\subsection{Split Computing} \label{split_computing}

\begin{figure}[t]
        \centerline{\includegraphics{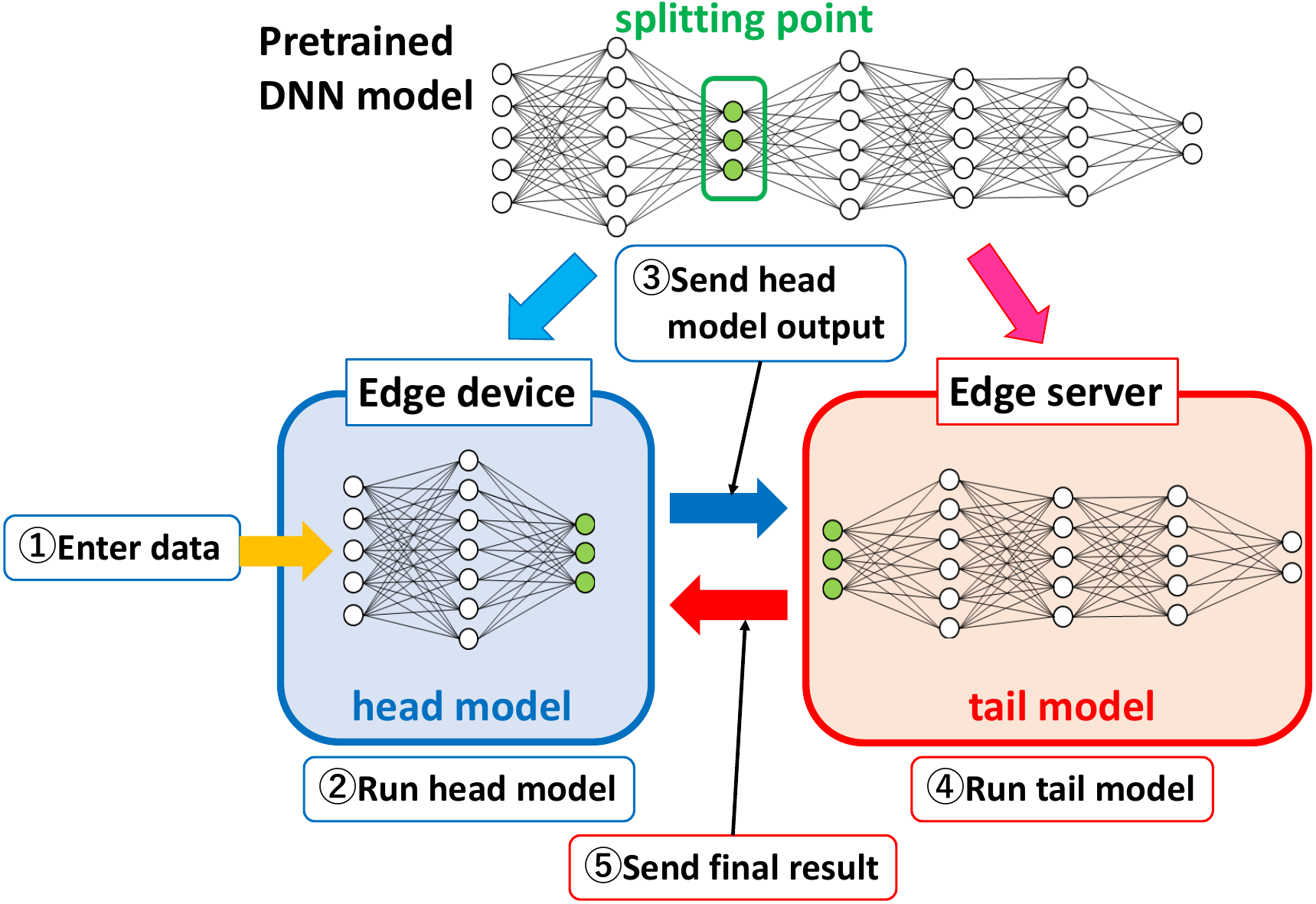}}
        \caption{Split Computing.}
    \label{fig-split_computing}
    \vspace{-2mm}
\end{figure}

In embedded systems, such as autonomous driving systems, the use of DNNs is typically performed on edge devices.
However, state-of-the-art DNN models are becoming increasingly complex in structure.
Because edge devices have limited computing power, most edge devices have difficulty meeting the high computational requirements of state-of-the-art DNN models.
Therefore, significant increases in processing time and power consumption can be problematic.
To address this problem, lightweight models with simplified model structures~\cite{sandler2018mobilenetv2, tan2019mnasnet} can be employed, but lightweight models usually suffer from reduced accuracy compared to state-of-the-art models.
Another approach is to send the data input to the edge device directly to the edge server and run the DNN model on the edge server.
However, this method requires a large amount of data transmission, which increases the communication time and processing time.
Moreover, given that the data input to the edge device contains privacy information, concerns arise regarding the risk of data interception and leakage.
One method that can address these issues is SC.

SC is a method in which the DNN model is split into two parts, and each part is executed on a different device.
This reduces the computational load on edge devices while maintaining high accuracy and processing speed.
Specifically, as shown in Fig.~\ref{fig-split_computing}, the pre-trained DNN model is first split into two parts in the middle.
The model on the input side is called the head model, and the model on the output side is called the tail model.
The head model is assigned to the edge device, while the tail model is assigned to the edge server.
During inference, the following steps are taken:
\begin{enumerate}
    \item The edge device receives the data.
    \item The edge device runs the head model.
    \item The output of the head model is sent to the edge server.
    \item The edge server runs the tail model using the received data as input and generates the final prediction result.
    \item The edge server sends this result to the edge device.
\end{enumerate}

SC reduces the amount of processing performed by edge devices.
Therefore, lightweight models do not need to be employed.
Processing can be optimized while maintaining high accuracy.
Setting the splitting point as early in the process as possible is also important.
This reduces the amount of computation on the edge device and saves power consumption.
Simultaneously, the processing time is significantly reduced since the majority of the model is executed on the edge server.

\subsection{OpenPCDet} \label{openpcdet}

\begin{figure}[t]
        \centerline{\includegraphics{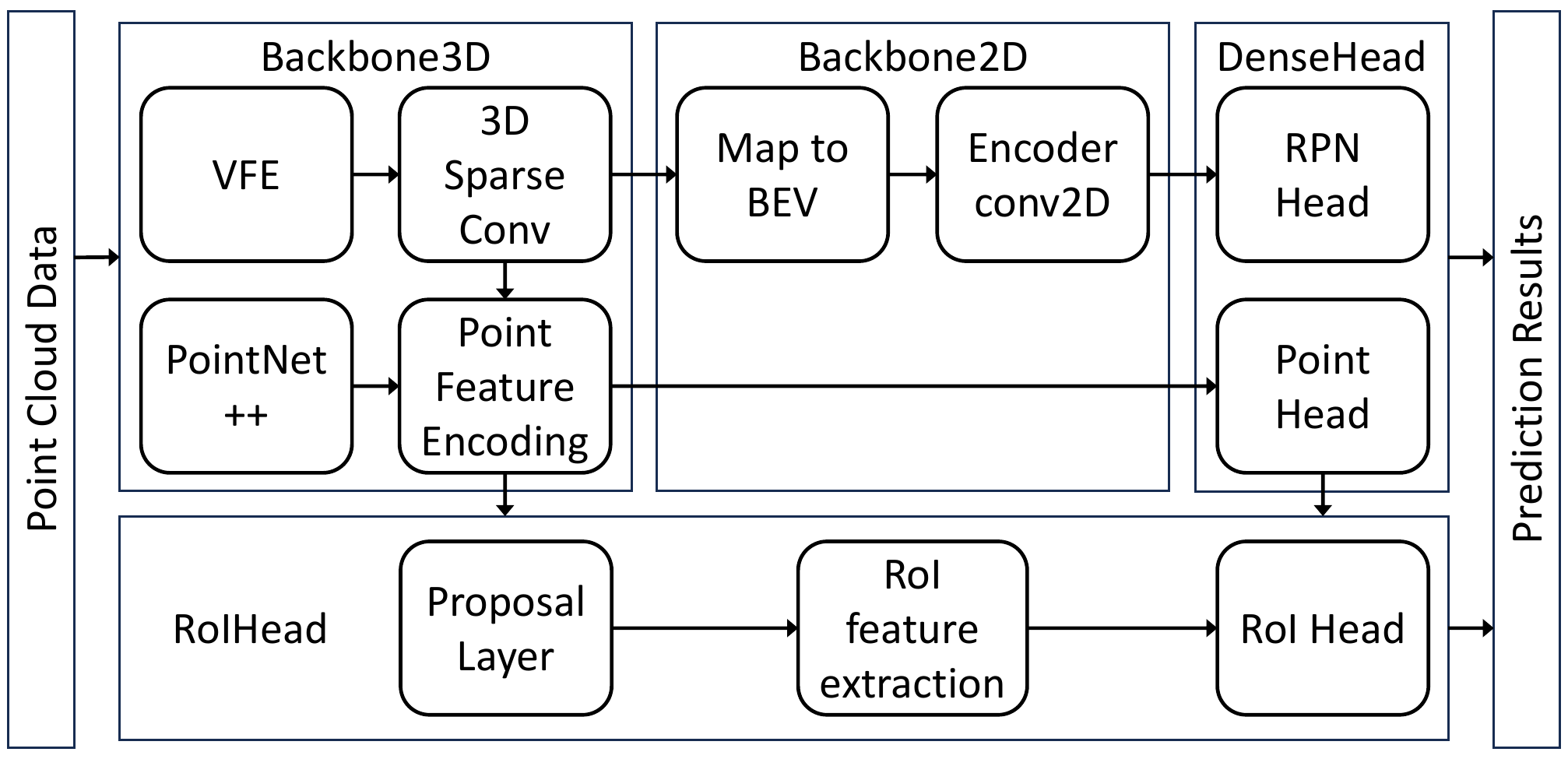}}
        \caption{The Base Module Structure of OpenPCDet.}
    \label{fig-OpenPCDet}
    \vspace{-2mm}
\end{figure}

OpenPCDet~\cite{openpcdet2020} is an open-source toolbox for LiDAR-based 3D object detection.
OpenPCDet supports a variety of 3D detection models and can be both trained and tested on each model.
To support different models, OpenPCDet has a generalized module structure.
The base module structure of OpenPCDet is shown in Fig.~\ref{fig-OpenPCDet}.
When OpenPCDet is run, the model is first constructed.
At that time, modules are constructed for each model setting according to the base module structure. 
The constructed modules are stored in order in a list.
During inference, the input point cloud data is first preprocessed.
Then, the constructed modules are executed in order.
Finally, the inference results are obtained from the module execution results.

\section{APPROACH} \label{approach}

\begin{figure}[t]
        \centerline{\includegraphics{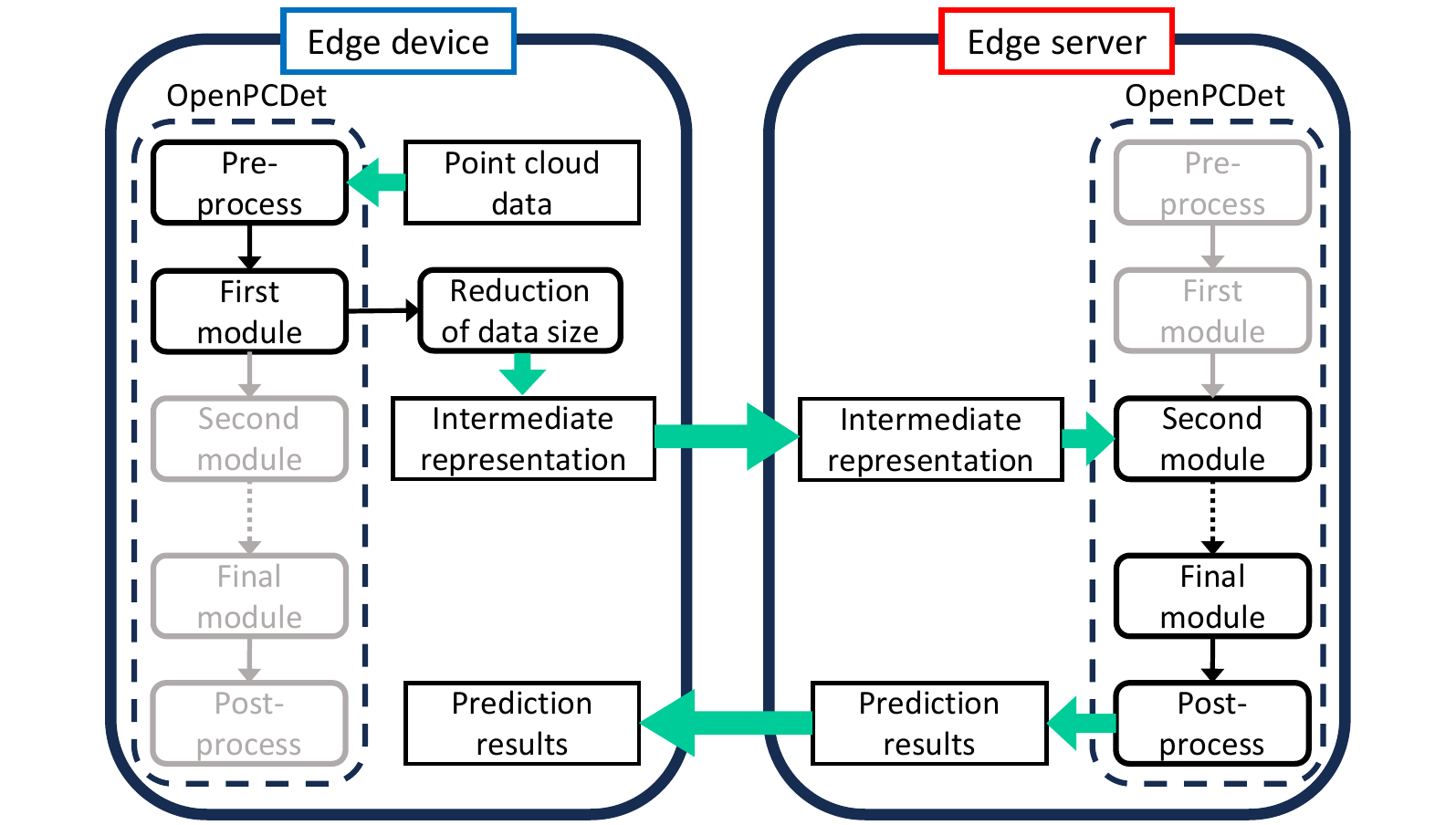}}
        \caption{Proposal Framework.}
    \label{fig-proposal_framework}
    \vspace{-2mm}
\end{figure}

In this study, the proposed method reduces the processing time in 3D object detection and the computational load on edge devices.
The proposed method is shown in Fig.~\ref{fig-proposal_framework}.
This method uses Split Computing for 3D object detection.

The remainder of this section is organized as follows.
Section~\ref{proposed_method} describes the proposed method of 3D object detection using Split Computing.
Section~\ref{two_approach} discusses the main approach to realizing the proposed method.

\subsection{3D object detection using Split Computing} \label{proposed_method}

This study proposes an object detection method that employs Split Computing to reduce the computational load on edge devices and the time required for object detection.
A concrete implementation of the proposed method is shown in Fig.~\ref{fig-proposal_framework}.
In the implementation, OpenPCDet is used for object detection.
For object detection, splitting points are defined in advance.
The edge device takes the point cloud data as input and executes the part of the model up to the splitting point.
After that, an intermediate representation with reduced data size is sent to the edge server.
The edge server executes the portion of the model after the splitting point and obtains the prediction results.
These prediction results are sent back to the edge device.

The proposed method has three advantages.
The first advantage is the reduction of computational load on edge devices.
In the conventional approach, the entire DNN model is executed on edge devices for object detection in infrastructure sensors.
However, the proposed method splits the DNN model into two parts, and only the head model on the input side is executed on the edge device.
This reduces the amount of computation on the edge device.

The second advantage is the reduction in overall object detection time.
Since edge devices have limited computing power, they can take a long time to execute a DNN model with a complex structure.
However, the proposed method executes part of the DNN model on the edge device and processes the remaining part on the edge server, which has more computing power.
This reduces the execution time of the DNN model compared to the case where the model is executed only on the edge device.
However, since the proposed method requires data transfer, reducing the data transfer time as much as possible is important.

The third advantage is that the point cloud data received from LiDAR is not sent as is.
To reduce the computational load on edge devices and the time required for object detection, one approach is to run the DNN model on the edge server.
However, this approach transmits the point cloud data received from LiDAR as is, and thus risks the leakage of privacy information contained in the point cloud data.
On the other hand, the proposed approach divides the DNN model in the middle and executes the model. 
Therefore, the data sent from the edge device to the edge server are the data in the middle of the DNN model.
Since the point cloud data is not sent as is, the risk of privacy information leakage is reduced.

\subsection{The main approaches required for realization}
\label{two_approach}

A main approach exists for the proposed method.
The approach is the selection of appropriate splitting points.
Splitting the DNN model is an important factor that has a significant impact on the performance of object detection.
For better object detection, two considerations should be made in selecting the splitting points.
The first is to split at the early stages of the model.
By placing the splitting points at the splitting points early in the model, the size of the head model allocated to the edge equipment can be reduced.
This reduces the computational load on edge devices and helps reduce the overall time required for object detection.

The second is to split the model at the layer with the smallest output data size.
The proposed method requires the output of the head model to be sent from the edge device to the edge server.
Reducing the time required for this data transfer contributes to reducing the computational load on the edge devices and the overall time required for object detection.
To reduce the data transfer time, the size of the transferred data must be reduced.
Since the size of the transferred data depends on the output data size of the head model, the model must be split into layers with a smaller output data size.

\section{EVALUATION} \label{evaluation}

This section evaluates 3D object detection using Split Computing.
In this study, Jetson Orin Nano, hexa-core Arm 1.5 GHz CPU and mounting 1024-core NVIDIA Ampere architecture GPU with 32 Tensor Cores operating at 625 MHz and 8 GB of 128-bit LPDDR5 memory was used as the edge device.
The KITTI dataset~\cite{geiger2012we} was used as the dataset.

The remainder of this section is organized as follows.
Section~\ref{selection_model} describes the selection of models for 3D object detection.
Section~\ref{eval_SC} compares object detection using Split Computing with object detection using only edge devices.

\subsection{Selection of models for 3D object detection} 
\label{selection_model}

\begin{figure}[t]
        \centerline{\includegraphics{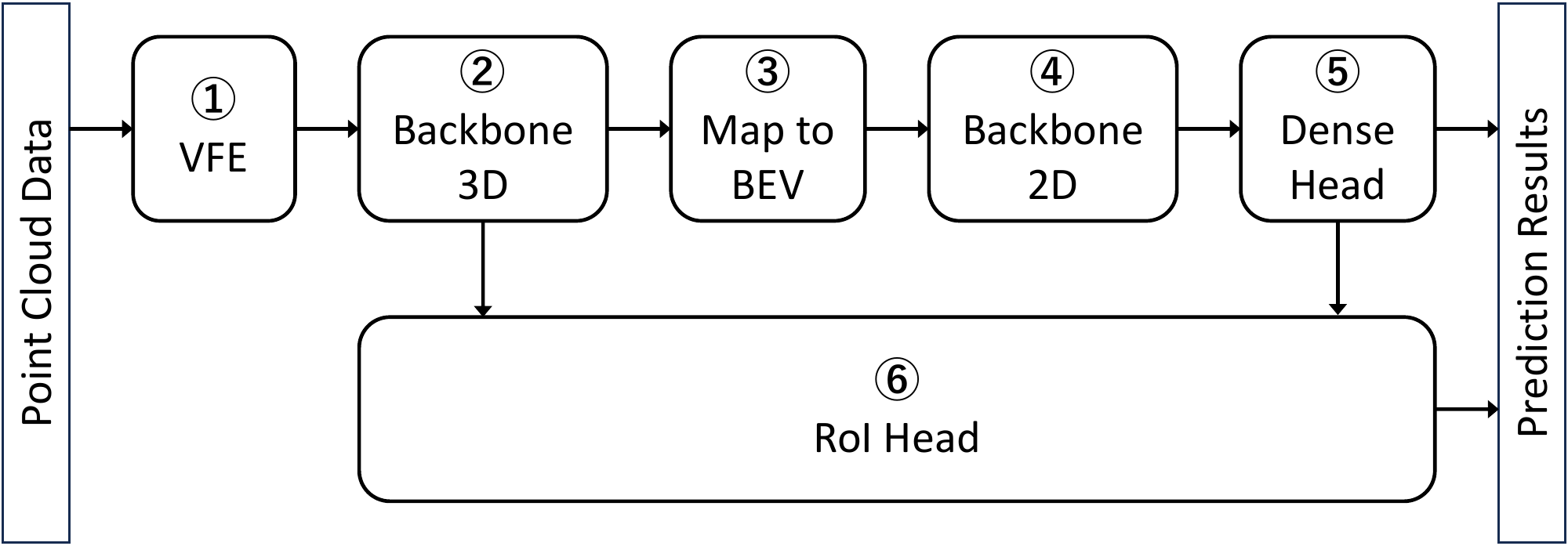}}
        \caption{The Module Structure of Voxel R-CNN in OpenPCDet.}
    \label{fig-VoxelRCNN}
    \vspace{-2mm}
\end{figure}

\begin{table}[t]
    \centering
    \caption{The ratios of the execution time of each module in Voxel R-CNN to the total}
    \begin{tabular}{|c|c|>{\raggedleft}p{25mm}c|}
         \hline
         \multirow{2}{*}{execution order} & \multirow{2}{*}{module} & \multicolumn{2}{c|}{ratio of module execution time}  \\
         \ & \ & \multicolumn{2}{c|}{to total time} \\ \hline
         \textcircled{\scriptsize 1} & VFE & 0.16869\% & \ \\ \hline
         \textcircled{\scriptsize 2} & Backbone 3D & 33.55415\% & \ \\ \hline
         \textcircled{\scriptsize 3} & Map to BEV & 0.28388\% & \ \\ \hline
         \textcircled{\scriptsize 4} & Backbone 2D & 2.43162\% & \ \\ \hline
         \textcircled{\scriptsize 5} & Dense Head & 1.15625\% & \ \\ \hline
         \textcircled{\scriptsize 6} & RoI Head & 62.40541\% & \ \\ \hline
    \end{tabular}
    \label{tab-ratio_module_time_VoxelRCNN}
\end{table}

When using Split Computing, the model that only uses the input point cloud data in the first step must be selected.
In this study, the model for 3D object detection used is Voxel R-CNN~\cite{deng2021voxel}.
The module structure and the execution order of Voxel R-CNN in OpenPCDet are shown in Fig.~\ref{fig-VoxelRCNN}.
Voxel R-CNN consists of six modules, and the input point cloud data is only required in the first module.
To define splitting points, the ratios of the execution time of each module in Voxel R-CNN to the total were measured.
The results are shown in Table~\ref{tab-ratio_module_time_VoxelRCNN}.
In Voxel R-CNN, the second module ``Backbone 3D'' and the sixth module ``RoI Head'' occupy the majority of the module's processing.
Therefore, splitting before the second module``Backbone 3D'' is appropriate.
Specifically, two patterns are used: splitting between the pre-process and the first module ``VFE,'' and splitting between \textcircled{\scriptsize 1}VFE and \textcircled{\scriptsize 2}Backbone 3D.

\subsection{Evaluation of 3D Object Detection Using Split Computing}
\label{eval_SC}

\begin{table}[t]
    \centering
    \caption{Elements of the transfer data for each splitting pattern}
    \begin{tabular}{|c|c|c|c|c|}
         \hline
         Splitting pattern & Conv1 & Conv2 & Conv3 & Conv4 \\ \hline
         Convolution layer & \multirow{3}{*}{Conv1} & \multirow{3}{*}{Conv2} & \multirow{2}{*}{Conv2} & Conv2 \\ 
         whose outputs are required & \ & \ & \multirow{2}{*}{Conv3} & Conv3 \\
         for elements of the transfer data & \ & \ & \ & Conv4 \\ \hline
    \end{tabular}
    \label{tab-element_data}
\end{table}

The proposed method, 3D object detection using Split Computing, was compared with the case where 3D object detection was performed using only edge devices.
The model for 3D object detection used here is Voxel R-CNN.
The split points are established after VFE, where voxelization is performed, and in Backbone3D, the network that runs after VFE.
Backbone 3D consists of four convolutional layers.
The convolution layers are 1x, 2x, 4x, and 8x in order from the input side.
Here, as shown in Fig.~\ref{fig-VoxelRCNN}, the output data of Backbone 3D is not only input to Map to BEV, the module that is executed immediately after, but also to RoI Head at the end of the process.
RoI Head uses the output of the second, third, and fourth convolution layers of Backbone 3D.
Therefore, as shown in Table~\ref{tab-element_data}, when splitting after the third convolutional layer, the outputs of the second and third convolutional layers need to be transferred from edge devices to edge servers.
Similarly, if splitting after the fourth convolutional layer, the outputs of the second, third, and fourth convolutional layers need to be forwarded from edge devices to edge servers.
For this reason, the performance of splitting after the third and fourth convolutional layers is clearly inferior to that of splitting after the second convolutional layer.
This is due to the increased computation at edge devices with less computing power, as well as the larger transfer data size.
Considering this model structure, in this study, the splitting points are set at two locations after the first and second convolutional layers, and each splitting pattern is compared.

\begin{figure}[t]
        \centerline{\includegraphics{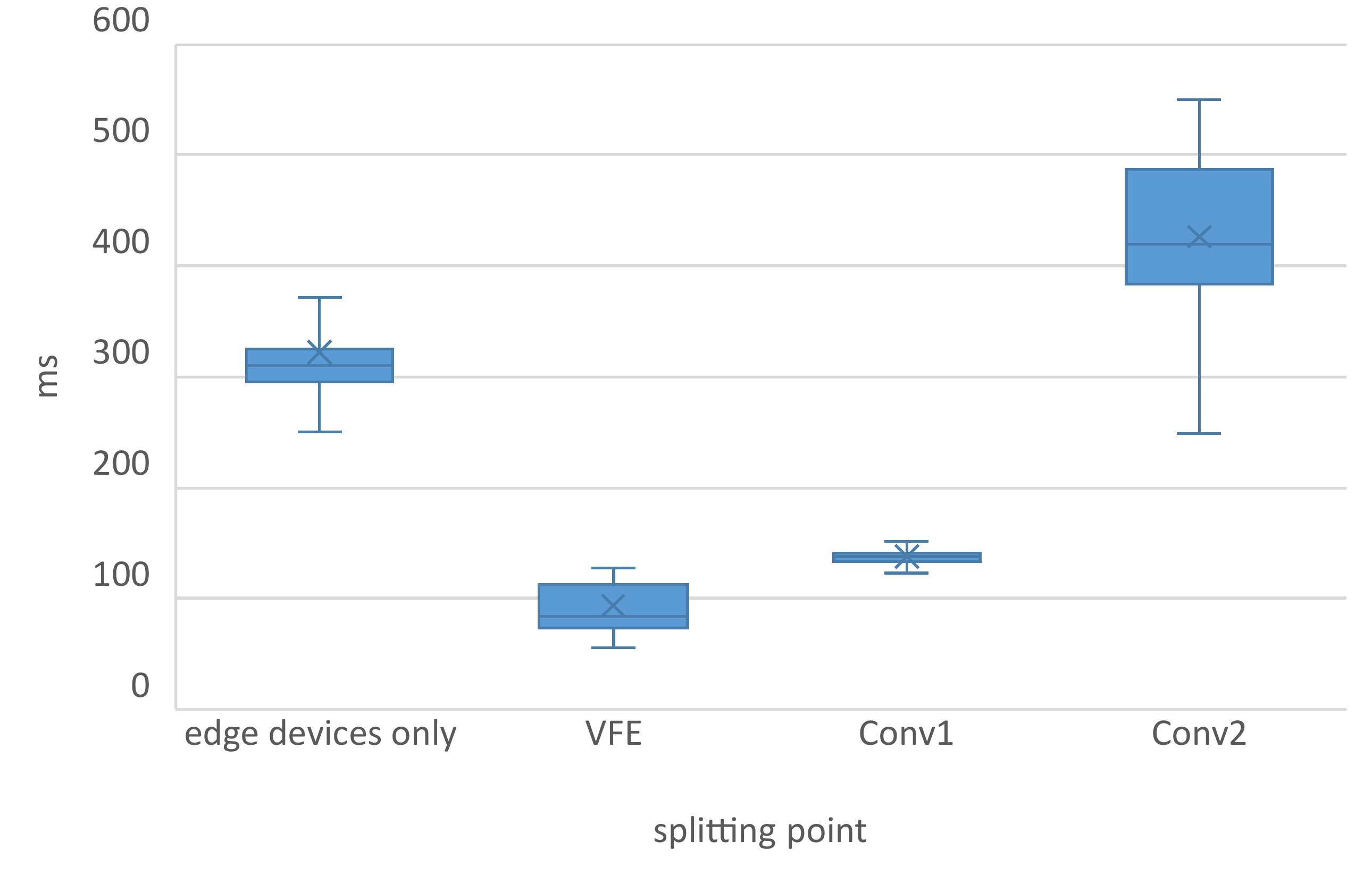}}
        \caption{Inference Time for Object Detection When Split within Backbone 3D.}
    \label{fig-inference_time_B3}
    \vspace{-2mm}
\end{figure}

First, the comparison is made in terms of the inference time.
The inference time is shown in Fig.~\ref{fig-inference_time_B3}.
When object detection was performed using only edge devices, the average inference time was 322 ms per scene.
On the other hand, the average inference time for the proposed method was 93.9 ms when split after VFE, 138 ms when split after the first convolutional layer, and 426 ms when split after the second convolutional layer.
In other words, the inference time was reduced by 70.8\% when split after VFE and by 57.1\% when split after the first convolutional layer, compared to object detection using only edge devices.

\begin{figure}[t]
        \centerline{\includegraphics{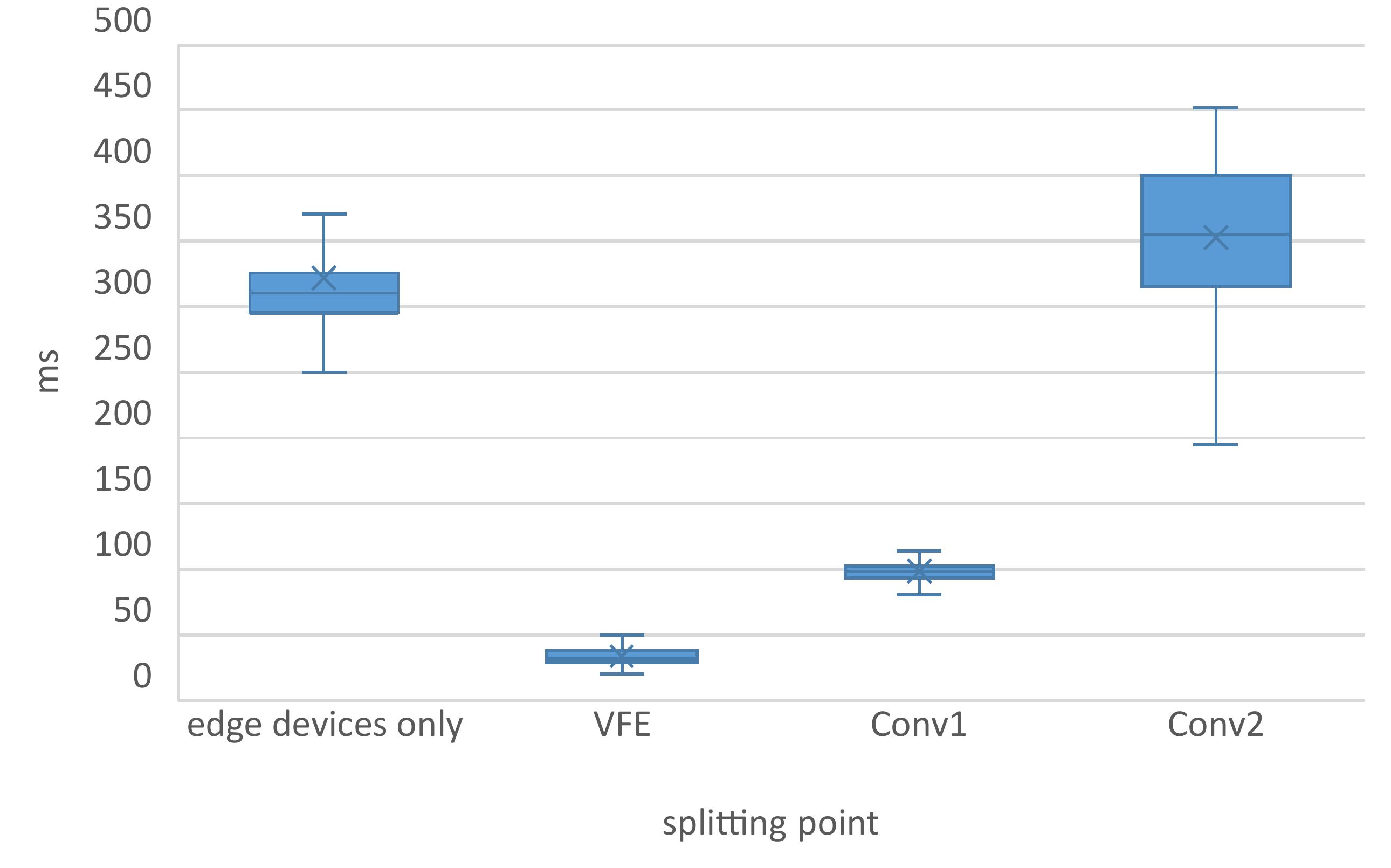}}
        \caption{Execution Time of the Edge Device from the Start of Inference to the End of Data Transfer to the Edge Server When Split within Backbone 3D.}
    \label{fig-edge_time_B3}
    \vspace{-2mm}
\end{figure}

Next, the execution time of edge devices, which is from the start of the inference to the end of data transfer to edge servers, is compared.
The execution time of edge devices is shown in Fig.~\ref{fig-edge_time_B3}.
When object detection is performed using only edge devices, the execution time of the edge devices is equal to the inference time.
In other words, the average execution time of the edge device is 322 ms.
On the other hand, the average inference time for the proposed method was 33.6 ms when split after VFE, 98.2 ms when split after the first convolutional layer, and 353 ms when split after the second convolutional layer.
In other words, the edge device execution time was reduced by 90.0\% when split after VFE and by 69.5\% when split after the first convolutional layer, compared to object detection using only edge devices.

\begin{figure}[t]
        \centerline{\includegraphics{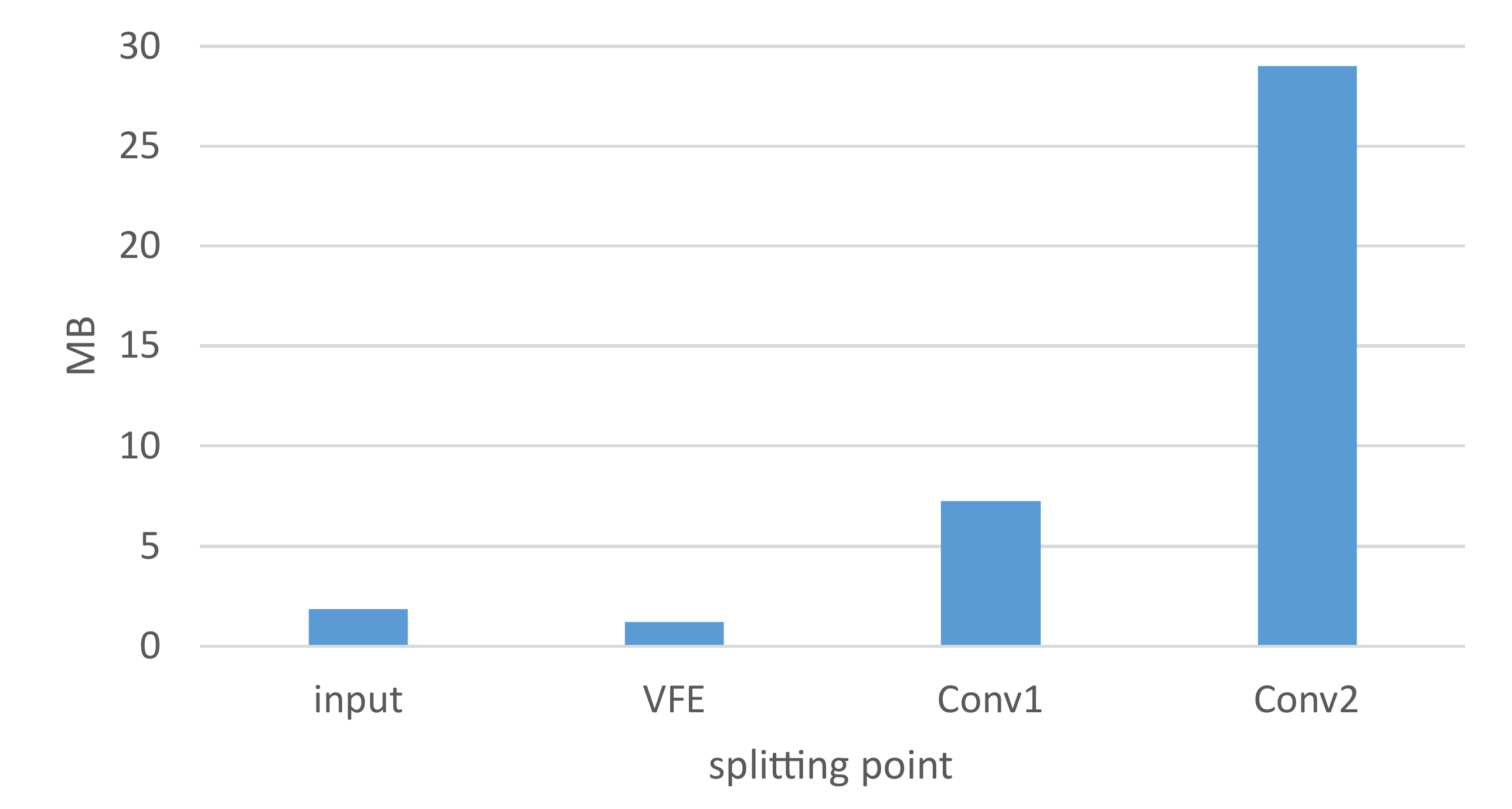}}
        \caption{Average Size of the Transfer Data for Each Splitting Pattern When Split within Backbone 3D.}
    \label{fig-send_datasize_B3}
    \vspace{-2mm}
\end{figure}

\begin{figure}[t]
        \centerline{\includegraphics{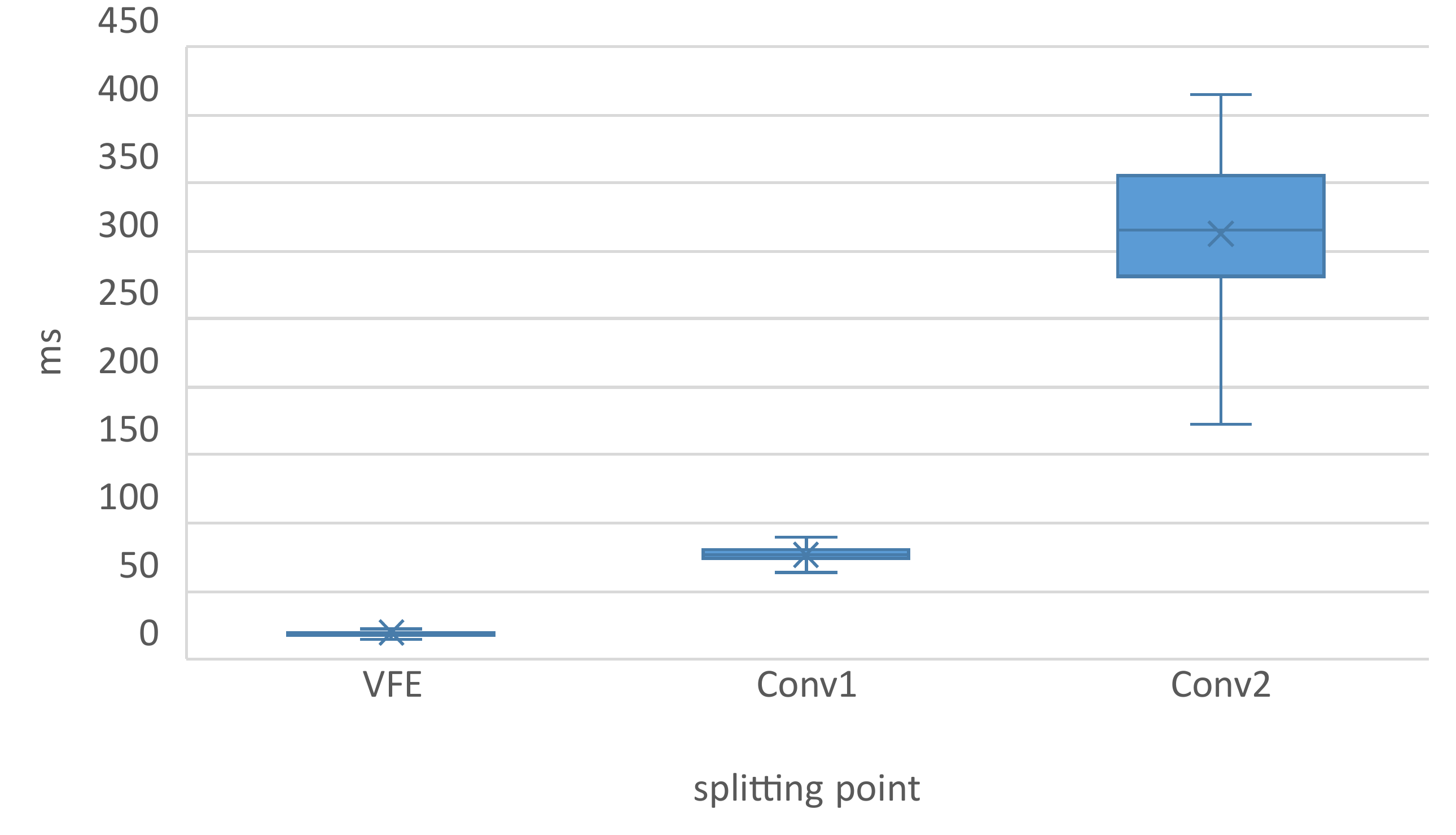}}
        \caption{Data Transfer Time for Each Splitting Pattern When Split within Backbone 3D.}
    \label{fig-send_time_B3}
    \vspace{-2mm}
\end{figure}

Finally, a comparison is made for data transfer.
First, the average size of the transfer data is shown in Fig.~\ref{fig-send_datasize_B3}.
The average data size of the input point cloud data was 1.84 MB.
On the other hand, the average transfer data size for the proposed method was 1.18 MB when split after VFE, 7.23 MB when split after the first convolutional layer, and 29.0 MB when split after the second convolutional layer.
Next, the data transfer time is shown in Fig.~\ref{fig-send_time_B3}.
The average data transfer time was 19.2 ms when split after VFE, 77.0 ms when split after the first convolutional layer, and 313 ms when split after the second convolutional layer.
Only when split after voxelization, the size of the data was smaller than transferring the point cloud data as is.
On the other hand, when split within the network, the size of the transfer data increased significantly.
However, point cloud data contains privacy information.
Similarly, privacy information is possible to be obtained from voxel data.
Therefore, considering the privacy risk, the data in the middle of convolution should be transferred even if the data size is large.
In fact, both the inference time and the edge device execution time are reduced when split after the first convolution.
Thus, if the privacy risk is to be considered, splitting within the network instead of splitting after voxelization is considered, even if the inference time increases a little.

\section{RELATED WORK} \label{related_work}

\begin{table}[t]
    \centering
    \caption{Comparison of the proposed method with related studies}
    \scalebox{1.0}{
    \begin{tabular}{|l|c|c|c|} \hline
        \ & Edge & Split & 3D Object \\
        \ & Device & Computing & Detection \\ \hline
        BottleFit~\cite{bottlefit} & \checkmark & \checkmark & \ \\ \hline
        Neural Rate Estimator and & \multirow{3}{*}{\checkmark} & \multirow{3}{*}{\checkmark} & \ \\
        Unsupervised Learning & \ & \ & \ \\
        in Split-DNN models~\cite{ahuja2023neural} & \ & \ & \ \\ \hline
        Voxel R-CNN~\cite{deng2021voxel} & \ & \ & \checkmark \\ \hline
        M3DeTR~\cite{guan2022m3detr} & \ & \ & \checkmark \\ \hline
        A Lightweight Model for & \multirow{3}{*}{\checkmark} & \ & \multirow{3}{*}{\checkmark} \\
        3D Point Cloud & \ & \ & \ \\
        Object Detection~\cite{li2023lightweight} & \ & \ & \ \\ \hline
        Proposed method & \checkmark & \checkmark & \checkmark \\ \hline
    \end{tabular}
    }
    \label{comparison}
\end{table}

Various studies have been conducted on SC and object detection models.
This section presents related works on the topic of SC and object detection models.
Along with the introduction, the proposed method is compared with related studies.
The results of the comparison are shown in Table~\ref{comparison}.

BottleFit~\cite{bottlefit} is an SC approach that introduces a bottleneck in the DNN model and devises a training method.
BottleFit modified the original model using a method that reduces the number of layers while still generating compressed layers.
As a result, bottlenecks were introduced without increasing the complexity of the model.
Furthermore, the learning phase involved a two-step process of learning the modified head model and learning the tail model to adjust to the modified head model.
This allowed BottleFit to achieve higher accuracy while reducing communication latency and power consumption compared to previous methods.

Neural Rate Estimator and Unsupervised Learning for Efficient Distributed Image Analytics in Split-DNN models~\cite{ahuja2023neural} is also part of SC's approach to introducing a bottleneck in the DNN model.
In this paper, when training the bottleneck, only the bottleneck weights were trained and the original DNN parameters were used without modification.
In addition, unsupervised learning with knowledge distillation was introduced because of the difficulty of collecting and labeling a large amount of training data.
This achieved a higher compression ratio than conventional compression methods, while maintaining the same level of accuracy.

Voxel R-CNN~\cite{deng2021voxel} is a 3D object detection model that takes point cloud data as input.
In this model, the point cloud is voxelized to reduce the computational load.
Features are extracted from the voxels using 3D Backbone Network.
Next, the voxel features are converted to 2D images, and the 2D Backbone Network and RPN are applied to generate 3D region proposals.
Finally, for each proposed region using Voxel RoI Pooling, RoI features are generated from the voxel features in that region.
This allows for faster processing while maintaining accuracy comparable to point cloud-based methods.

M3DeTR~\cite{guan2022m3detr} is also a 3D object detection model that takes point cloud data as input.
In 3D object detection, the three main representations used to process point cloud data are voxels, raw point clouds, and bird's-eye views.
Each representation has unique advantages and can be combined to improve detection accuracy.
M3DeTR introduces M3 Transformer to aggregate multiple representations, multi-scale representations, and the interrelationships of points in point cloud data.
M3 Transformer uses the Transformer architecture and leverages a self-attention mechanism to model dependencies and learn rich information.
With this implementation, M3DeTR has achieved higher accuracy than previous studies.

A Lightweight Model for 3D Point Cloud Object Detection~\cite{li2023lightweight} is a lightweight model for 3D object detection using point cloud data.
The goal of this paper is to explore models that require only computational requirements that can be met by edge devices with limited computational power.
In this paper, LW-Sconv module is proposed, which can reduce the complexity of the model.
The accuracy of the model is also improved by employing knowledge distillation.
This allows for reduced computational cost with little loss of detection accuracy.

\section{CONCLUSION} \label{conclusion}

This paper proposed the Split Computing method for 3D point cloud object detection on edge devices with limited computational power.
OpenPCDet, a LiDAR-based open-source toolbox, was used for 3D object detection.
To implement Split Computing method, a split point was defined in the model.
Then, modules before the split point was executed on edge devices and modules after the split point was executed on edge servers.

In this study, the Jetson Orin Nano was employed as the edge device to evaluate the proposed method.
Object detection using only the edge device and Split Computing method were compared in terms of inference time and edge device execution time.
The results showed that Split Computing reduced the inference time and the computational load on the edge device.

In this study, the output of the head model was transferred as is.
However, by compressing the transfer data using quantization or other methods, the transfer data size is reduced, and the transfer time is shortened.
Currently, the system is designed to process point cloud data from a single LiDAR unit.
However, the system becomes more practical by building a system that is capable of processing integrated data from multiple LiDARs.

\section*{Acknowledgment} 
This work was supported by JSPS KAKENHI Grant Number 23H00464.

\bibliographystyle{IEEEtran}
\bibliography{reference}

\end{document}